\documentclass[conference]{IEEEtran}
\usepackage{newtxtext}
\IEEEoverridecommandlockouts

\usepackage{cite}
\usepackage{amsmath,amssymb,amsfonts}
\usepackage{algorithmic}
\usepackage{graphicx}
\usepackage{textcomp}
\usepackage{xcolor}
\usepackage{balance}
\def\BibTeX{{\rm B\kern-.05em{\sc i\kern-.025em b}\kern-.08em
    T\kern-.1667em\lower.7ex\hbox{E}\kern-.125emX}}
\begin{document}

\title{Monitoring Continuous Integration Practices in Industry: A Case Study\\
}

\author{\IEEEauthorblockN{Jadson Santos}
\IEEEauthorblockA{\textit{Informatics \& Applied Math Dept.} \\
\textit{Federal University of Rio Grande do Norte}\\
Natal, Brazil \\
jadson.santos@ufrn.br}
\and
\IEEEauthorblockN{Daniel Alencar da Costa}
\IEEEauthorblockA{\textit{School of Computing} \\
\textit{University of Otago}\\
Dunedin, New Zealand \\
danielcalencar@otago.ac.nz}
\and
\IEEEauthorblockN{Uirá Kulesza}
\IEEEauthorblockA{\textit{Informatics \& Applied Math Dept.} \\
\textit{Federal University of Rio Grande do Norte}\\
Natal, Brazil \\
uira@dimap.ufrn.br}
}

\maketitle

\begin{abstract}
  In this paper, we study the benefits and challenges of monitoring Continuous Integration (CI) practices in software development. Our aim is to evaluate the impact of monitoring seven CI practices in industry using three organizations in Brazil as case studies. We developed a tool for monitoring CI practices and conducted a multiple case study, applying a mixed-methods strategy. We combined surveys, interviews, log data, and repositories data from software projects and their CI services.We gauged the organization's interest in monitoring CI practices. The act of monitoring CI provided an overview of the organizational state of practice in terms of CI, motivated further improvement of CI practices, increased perceived quality of software, and improved project communication. We recommend that companies adopt the practice monitoring of CI practices and that CI services integrate monitoring functionalities into their dashboards.
\end{abstract}

\begin{IEEEkeywords}
CI Practices, CI Metrics, Monitoring, Case Study, CI Maturity
\end{IEEEkeywords}

\section{Introduction}

Continuous Integration (CI) is a software development practice in which team members frequently integrate their work \cite{FowlerCI2006}. The integration process involves an automated build step that includes compiling the code and running automated tests. This ensures the detection of integration errors as quickly as possible. Despite extensive research on the benefits and costs of Continuous Integration \cite{StakeholderPerceptions2015, ReplicationCIPainPoints2019, TradeOffsCI2017, ImpactCICodeReviews2020, hilton2016usage, benefitsCIIndustry2013, elazhary2021uncovering, VasilescuQualityProdOutcomes2015, JoaoHelis2018, ghaleb2019empirical, coverageImpac2017, JadsonESEM}, most studies focused on mining repositories to investigate the overall impact of CI. Although understanding the impact of CI is important, as a software engineering community, we do not know (1) whether development teams monitor their CI practices; (2) whether there exists willingness to monitor CI practices; and (3) which benefits the act of monitoring practices potentially bring to the development process. The focus of our research is to address these gaps.

In their systematic review of the literature (SLR), Eliezio et al. \cite{EliezioRevisaoSistematica2022} identified a lack of criteria to determine whether projects under study use CI properly. They found that 42.5\% of the primary studies did not apply or establish any criteria to determine whether a project uses CI properly. Among those which established some criteria, 56.25\% of them relied on a single criterion. The most common criterion applied was the adoption of a CI service, such as Travis CI. In a similar vein, other studies \cite{CITheater2019} have demonstrated that several projects using CI services do not employ all CI practices \cite{JadsonESEM}. For example, 60\% of the projects practiced infrequent commits; (ii) 85\% of the projects contained at least one broken build which took a long time to be fixed; and (iii) most of the projects contained builds which lasted more than 10 minutes to be generated. 

Considering that projects may have difficulties to implement the best practices of CI, our paper investigates whether the act of monitoring CI practices can foster a better CI environment in industry. To achieve this goal, we conducted a multiple case study with mixed methods, combining survey, interviews and access log analysis. For each project, we applied an initial survey to understand the level of CI maturity and the team's satisfaction with CI. This allowed us to better compare and understand the impact of monitoring CI practices on projects at the end of the study. Additionally, we developed a tool \cite{JadsonSBES, JadsonESEM} with an extensive suite of metrics that automatically calculates the values of seven CI metrics (related to CI practices we aim to monitor), enabling developers to continuously access these metrics and receive weekly alerts about their evolution.

Over a period of 8 weeks, we conducted a series of interviews with team members to gain insights into the benefits and challenges of monitoring the investigated CI practices. We applied the suite of metrics in three software development projects from different Brazilian organizations. By collecting metrics from CI services, access log, surveys and interviewing organizational staff, we answered the following research questions:

\begin{itemize}
\item \textbf{RQ1: Are developers satisfied with the project's CI maturity levels?} By surveying and interviewing members of the project, we assess whether developers are satisfied with the current level of CI in their project. This exploratory research question provides insights into the need to monitor CI practices. Our findings indicate that $\frac{2}{3}$ of the projects were satisfied with the CI maturity level even before we conducted our case study. However, developers did not use any metric-driven assessment to evaluate the CI maturity at this stage. Despite initial satisfaction levels, CI monitoring helped participants notice several aspects of CI that required improvement.

\item \textbf{RQ2: What benefits does CI monitoring bring?} By interviewing project staff, we collected their perceptions regarding the benefits of monitoring CI practices. Our findings reveal that participants identified several potential benefits from monitoring CI, including the ability to obtain an overview of the current practice with respect to CI, to highlight problems related to CI, and serve as a reminder and motivator for improving and practicing CI, among other advantages.

\item \textbf{RQ3: What are the challenges/disadvantages of monitoring CI practices on a daily basis?} By interviewing project staff, we collected their perceptions of potential challenges and disadvantages related to monitoring CI practices. There were no significant challenges directly related to monitoring CI practices nor the use of our monitoring tool. External problems, such as deadline management, instability in the CI environment, and a lack of emphasis on the importance of CI, were the main challenges faced by projects.

\item \textbf{RQ4: Are developers interested in using our metrics suite tool?} By collecting the logs of access to the dashboard of our metrics suite tool (which contains the suite of metrics), and without the knowledge of the participants, we assessed how frequently the tool was used during the case study, thus measuring real interest in monitoring CI practices. The record of access to the dashboard screen allows us to conclude that there was considerable access to the tool during the case study, even outside the interview days. This was mainly the case for companies that already hold a certain level of CI maturity.

\item \textbf{RQ5: How did CI practices evolve during the use of our metrics suite tool?} By comparing the evolution of CI practices throughout the study period, we gain insights into the role of monitoring in this evolution. We noticed a notable enhancement in CI practices with one project, but for the other two projects, due to the instability of the CI environment and the break for the end-of-year holiday, we cannot perceive a constant improvement.

\end{itemize}

\section{Related Work}

In this section, we present an overview of research works related to monitoring of CI practices.

Wang et. al. \cite{wang2022test} conducted an empirical study to observe the impact of test automation maturity on product quality, test automation effort, and release cycles in the context of CI in open-source projects. They conducted a test automation maturity survey with developers from 149 open-source Java projects. ``Test automation maturity'' was quantified as the total score obtained in all maturity-related questions in their survey. Their findings indicated that higher levels of test automation maturity were positively correlated with improved product quality (p-value=0.000624) and shorter release cycles (p-value=0.01891). 

Sallin et. al. ~\cite{sallin2021measuring} conducted a comprehensive literature review to assess the feasibility of automatically measuring the four DORA Metrics \cite{DoraMetricsLeanIX2023}. They found that the scientific literature has not investigated the automatic measurement of DORA metrics. Their examination of grey literature revealed 16 articles discussing various aspects of automatic measurement. The authors also conducted a case study involving a Scrum team comprising ten individuals to evaluate the benefits of automating the measurement of DORA metrics. 

Elazhary et. al. \cite{elazhary2021uncovering} investigated the extent to which three software organizations implement ten Continuous Integration (CI) practices defined by \cite{FowlerCI2006}. They explored the benefits these practices bring and the challenges encountered during their implementation. This inquiry was conducted through a multiple case study with mixed methods, focusing on three small to medium-sized software-as-a-service organizations. The study revealed that four practices exhibited significant differences in adoption between organizations.


Differently from previous work, our research focuses on the monitoring of CI, which consists of broader practices, such as short build durations and test coverage. For example, although Wang et. al. \cite{wang2022test} focused on a single CI practice, specifically test automation, we evaluate the monitoring of a larger group of practices. In contrast with the Sallin et. al.'s ~\cite{sallin2021measuring} work, we focus on evaluating the advantages of monitoring CI practices, extending beyond the scope of sole automation. Perhaps the work most aligned with ours is the work proposed by Elazhary et. al.\cite{elazhary2021uncovering}. However, our analysis focuses on different CI practices. Moreover, different from all previous work, we developed a metrics suite tool \cite{JadsonSBES, JadsonESEM} that enabled us to perform case studies, collecting and visualizing the CI practices of three software development companies. The implemented tool suite automatically collects and delivers 7 CI metrics values to developers. This represents a significant advantage, particularly for the organization undergoing the case study. Our primary focus lies in assessing the benefits and challenges associated with monitoring CI practices over time. In contrast, Elazhary et. al. \cite{elazhary2021uncovering} emphasis was on exploring the benefits and challenges of implementing CI practices.

\section{Methodology}

We conducted a multiple case study \cite{stake2013multiple} on monitoring the CI practices of three software projects from three different organizations. The main inclusion criteria for organizations in this study were: (i) location – we opportunistically selected organizations where we could interact with developers and personally visit the organization yielding more confidence in our data analysis; (ii) availability – organizations had to provide a basic infrastructure to run our tool and allocate some of their developer time for interviews and validation of our findings; and (iii) the organization had to have at least one project that used a CI service and generating regular builds.

The characteristics of the selected organization are:

\begin{itemize}
    \item \textbf{Department of IT at a Public University (Organization A)}:
        Organization A is a public educational institution, specifically a department of Information Technology (IT) within a prominent public university in Brazil. It encompasses a wide range of academic and administrative IT projects, such as: a job vacancy system, a technology park management system, an access control system, a website, etc. The university is known for its commitment to research and technological advancements, making it an ideal candidate for studying CI practices.

    \item \textbf{Global Technology Company (Organization B)}:
        Organization B is a global technology company with a local development team based in our city, which allowed its inclusion in our case study. The software project monitored in our case study focuses on cutting-edge technology and innovation in 5G networks, offering a distinct perspective on CI practices from an industry standpoint.
        
    \item \textbf{State House of Representatives (Organization C)}:
        Organization C is a branch of the government of a Brazilian state, which is responsible for legislative activities. As a public institution, the State House of Representatives plays a vital role in shaping legislation and policies at the state level in Brazil. The development team works on developing the project's management software, such as: the human resources department system, website, among others.
\end{itemize}

Due to ease of access, Organization A was the first to initiate interviews of our case study, starting the interviews in November 2023 and concluding them in December 2023. We used this organization as a pilot for the case study, as recommended by best practices \cite{yin2009case}. Organizations B and C began their interviews in mid-December 2023 and completed them in early February 2024. We collected data related to CI practices from October 1, 2023, until the end of the interview period. This allowed us to track the evolution of practices over a period of three months for Organization A and four and a half months for Organizations B and C.

\begin{figure}[ht]
  \centering
  \includegraphics[width=\linewidth]{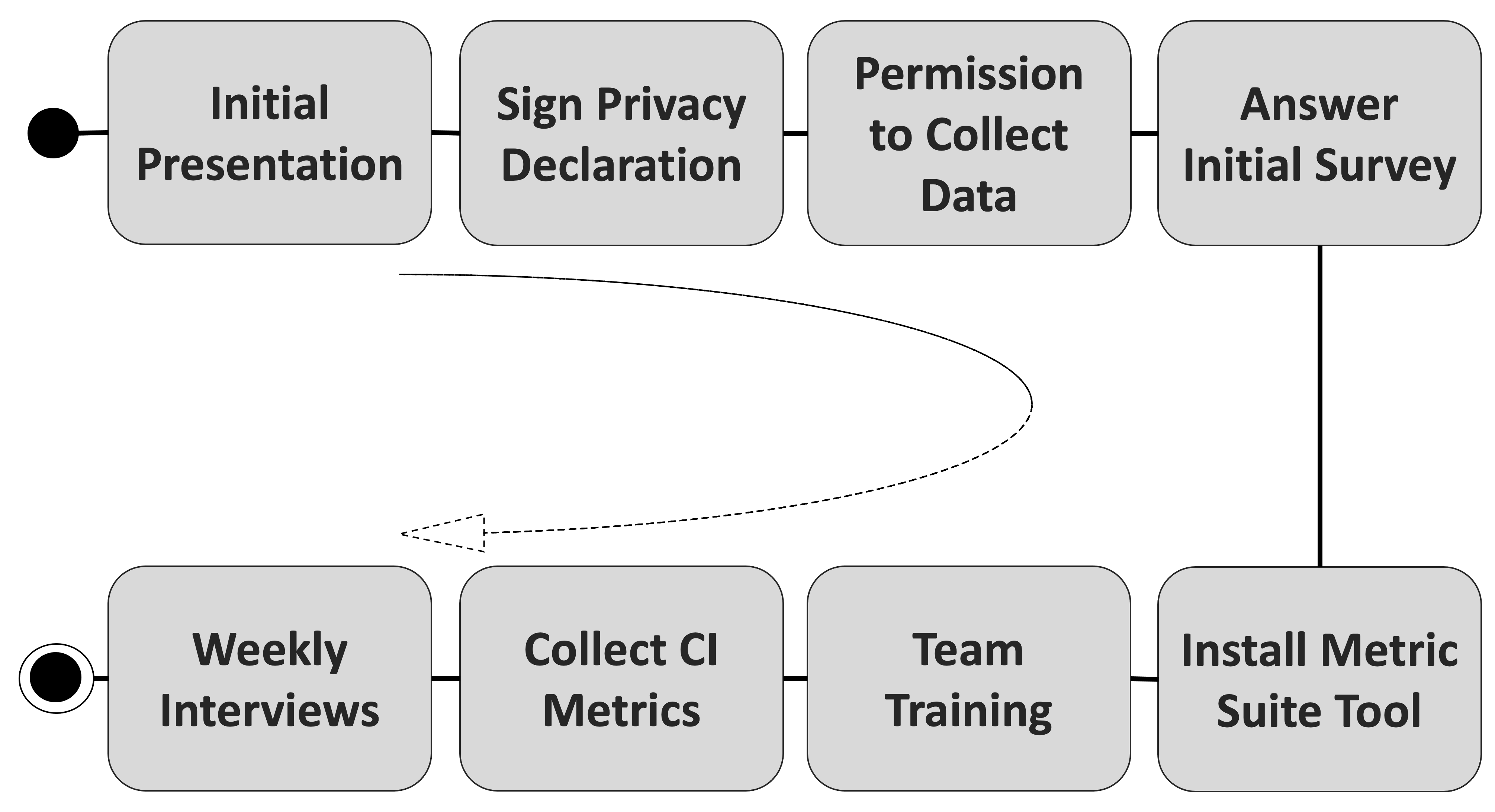}
  \caption{Data collection plan}
  \label{fig:data_collection_plan}
\end{figure}

Coincidentally, all three organizations participating in the case study utilized the same CI service: Gitlab. They employed Gitlab On-Premise, installed on the organizations' intranets, which required specific access requests for each entity. To gather metrics related to CI practices, we mined the build history of projects in the Gitlab CI service. Additionally, we collected code coverage metrics from the \textsc{Sonarqube} tool whenever available (Organizations C did not use Sonarqube). We selected key members of the development team as study participants. The criteria for selecting these members included development experience, basic knowledge of CI and familiarity with the company’s development environment. Figure~\ref{fig:data_collection_plan} provides an overview of all the steps involved in our data collection plan.

The data collection plan for the case study consists of an eight-step approach. It begins with an initial presentation to introduce the case study's objectives, importance, needs, and potential benefits for the participating projects. This step was important to convince the organizations to participate in the case study. Following the initial presentation, project representatives were asked to sign privacy declarations to ensure data confidentiality and to obtain permission to proceed with the study. We applied an initial survey (explained in detail in Section~\ref{subsec:survey}) to assess the current perceived maturity level of CI practices in each project. Subsequently, training sessions were conducted to familiarize the teams with the developed metric suite tool for CI monitoring. The tool was then installed on a server within the organization’s intranet, provided by the organization. This setup made it easier for participants to access the tool, which was available 24 hours a day in their work environment. The metric suite tool was then configured to start the weekly collection of CI practice data. Finally, we conducted weekly interviews with the development teams (explained in detail in Section~\ref{subsec:interview}) to capture real-time insights and measure the impact of monitoring CI practices. The CI metrics collected during our case study are described below:

\begin{itemize}

\item{\textbf{Commit Per Weekday/Commit Activity ~\cite{CITheater2019}}: Mean of the absolute number of commits according to the week day of the analyzed period.}

\item{\textbf{Coverage ~\cite{CITheater2019}}: Measures which parts of a program are executed when running the tests. Represents the percentage of the program exercised by tests.}

\item{\textbf{Build Duration ~\cite{CITheater2019}}: Measures the duration of the build (build finished at timestamp - build started at timestamp).}

\item{\textbf{Build Activity/Build-frequency ~\cite{STAHL201448}}: This metric is a unit interval (i.e., a closed interval [0,1]) that represents the rate of builds generated during the CI period. } 

\item{\textbf{Build Health/Build Quality ~\cite{EliezioRevisaoSistematica2022}}: This metric is a unit interval that represents the rate of build failures during the CI period. } 

\item{\textbf{Time to Fix a Broken Build ~\cite{CITheater2019}}: Measures the length of time (the mean time) that builds were left broken. When a build breaks, we computed the time in seconds until the build returned to the ``passed'' status.} 

\item{\textbf{Comments Per Change ~\cite{thompson2017large, benefitsCIIndustry2013}}: Measures the mean number of comments in a group of Merge Requests or Issues. This metric measures the level of communication between the team.}

\end{itemize}

In conducting the case study, we adhered to a systematic and structured methodology based on \cite{tellis1997application}. 1) The first phase involved the design of a comprehensive case study protocol. This protocol should determine the required skills for the case study and should be developed and reviewed. 2) The second phase centered on the actual execution of the case study, including adequate preparation for data collection, installation of the CI practices monitoring suite, and distribution of questionnaires to relevant participants within the selected projects to assess the perception of current CI maturity in the project. Subsequently, interviews were conducted weekly to gather qualitative insights focused on answering the study’s research questions. 3) The third phase focused on the meticulous analysis of the gathered evidence. 4) The fourth phase involved the synthesis of conclusions, recommendations, and implications based on the evidence obtained through the case study. We also adopted the best practices defined by \cite{yin2009case}: conduct a pilot case study (Organization A), train for a specific case study, screen candidate cases, ask good questions, be a good “listener,” stay adaptive, have a firm grasp of the issues being studied, and avoid biases by being sensitive to contrary evidence.

To encourage participation in the study, we also outlined the benefits for projects participating in the study. These benefits are: (i) the project would gain an overview of its CI maturity and processes, (ii) the monitoring suite tool, along with its source code, would be made permanently available and free of charge to the project, and (iii) at the end of the study, 10 additional metrics would be unlocked, including Continuous Delivery and DevOps metrics (DORA metrics\footnote{https://cloud.google.com/blog/products/devops-sre/using-the-four-keys-to-measure-your-devops-performance}).

\subsection{Survey} \label{subsec:survey}

To gain an understanding of the project's CI maturity before implementing the case study, we requested each developer to respond to a preliminary survey. This initial survey helped us understand the state of CI within each project prior to the case study, assess satisfaction with CI adoption, and identify any metrics used to measure CI. This preliminary insight facilitated a more comprehensive evaluation of the case study's impact on the evolution of CI in each analyzed project. Furthermore, the survey responses allowed us to answer RQ1.

The survey consisted of 15 questions, 9 of which were open-ended. To give respondents as much flexibility as possible, 9 questions were mandatory (6 of which were close-ended while 3 were open-ended), and 6 questions were optional (all open-ended). The survey was divided into four sections: 1) Introduction, 2) Personal Information, 3) Team Organization and Collaboration, and 4) Diagnosis of the project's CI maturity.

In the ``Introduction'' section of the survey, we provide context for the study, introduce the concept of CI, and outline the survey's objectives. We also specify the estimated time required for completion and assure participants that their responses will be used solely for research purposes and published anonymously. The ``Personal Information'' section assesses participants' knowledge of software development and Continuous Integration, ensuring they have the necessary qualifications for the case study. The ``Team Organization and Collaboration'' section explores the organization's CI environment, development processes, and tools. Finally, the ``Diagnosis of the Project's CI Maturity'' section investigates how the organization evaluates the CI maturity of its own projects, the metrics employed, and overall satisfaction with CI adoption.

\subsection{Interviews} \label{subsec:interview}

We conducted semi-structured interviews \cite{dearnley2005reflection} to obtain insights into the monitoring practices regarding the CI of the analyzed projects. We defined a set of initial questions, although these could vary depending on the participants' responses during the interviews. The interview questions were used primarily to address RQ2 and RQ3 and are detailed in the case study protocol. We followed the guidance provided by \cite{hancock2021doing}. We also adopted best practices before the interviews, including: 1) identifying key participants, 2) developing an interview guide/protocol, 3) ensuring a private and neutral environment, 4) recording the interview data, and 5) adhering to legal and ethical requirements. When conducting the interviews, we: 1) obtained consent and ensured anonymity, 2) reviewed the interview's purpose and duration, and 3) listened actively. We generated an alias for each participant to ensure anonymity. Table \ref{tab:case_participants} shows the study participants.

\begin{table}[ht]
    \caption{Case Study Participants}
    \centering
    \begin{tabular}{|c|l|c|}
        \hline
        \textbf{Label} & \textbf{Role} & \textbf{Project} \\
        \hline
        P1 & Software Developer & A \\ 
        \hline
        P2 & \begin{tabular}[c]{@{}l@{}}Software Developer; \\ Software Testing Analyst; \\ Application Architect\end{tabular} & B \\ 
        \hline
        P3 & Researcher & B \\ 
        \hline
        P4 & \begin{tabular}[c]{@{}l@{}} Software developer; \\ 5G networks Researcher \end{tabular}  & B \\
        \hline
        P5 & Software Developer & C \\
        \hline 
    \end{tabular}
    \label{tab:case_participants}
\end{table}

\subsection{Thematic Analysis of Interview Data}

Following a similar approach to \cite{elazhary2021uncovering}, we transcribed the interviews of all five participants. We assigned labels to participants' statements that sought to express, in a simple and summarized manner, the participants' perceptions regarding monitoring CI practices. We grouped common themes and created a network graph to represent them. From there, we conducted an analysis to extract the main findings from these themes. We shared each participant's final themes with them so they could review and indicate whether they agreed with the extracted themes.

\subsection{Study Replications} \label{subsec:study_replication}

We have made the documentation generated by this case study available to facilitate future replications at: \\ \textbf{https://zenodo.org/records/15507762}. We provide the use case protocol, the source code, as well as an example of the survey regarding the project's CI maturity. The CI metrics' values throughout the case study and interview transcriptions are included in an anonymous manner.

\section{Results}

\subsection{RQ1: Are developers satisfied with the project’s CI maturity levels?}


Organization A currently characterizes itself as operating at an INTERMEDIATE level of CI adoption. This classification did not involve the use of any specific CI practice. However, there is a recognized need for a strategic shift toward incorporating such practices into their CI process. The team believes that to begin adopting practices, it is necessary to plan which CI practices are appropriate for their context and determine how they would be implemented and used. When asked about their satisfaction with the current CI maturity level, they stated they were satisfied with the company's progress. Furthermore, there is a perception that adopting automated database migration techniques, eliminating manual database upgrade routines, and incorporating fully automated acceptance testing would significantly enhance the project's CI maturity.

Organization B identified itself as operating at an INTERMEDIATE level of CI maturity. The classification at this level is not determined by specific CI practices but rather by the successful automation of critical processes: the build process, automated testing, automatic generation of Docker images, and the deployment of new versions in a dedicated test/approval environment. Despite the absence of a metric-driven assessment, the participants expressed satisfaction with the current state of CI within the company. Nevertheless, they envisioned further improvements to enhance CI effectiveness. Participants believed that greater code coverage related to automated tests, incorporating tools focused on code security, and utilizing Flyway to automate database version evolution would contribute significantly to elevating the overall maturity and efficacy of the project's CI practices.

Organization C considered its current level of CI maturity as BASE level. This level lacks the incorporation of specific CI practices to precisely delineate the extent of adoption. Despite this, the case study participant expresses a notable dissatisfaction with the existing level of CI within the company, emphasizing that CI is still in its initial stages of implementation.

\begin{center}
\fbox{
    \begin{minipage}{24em}
    \textbf{ Our results reveal that, except for Organization C — which is still in the early stages of CI adoption and performs most processes manually — the other projects were satisfied with their CI maturity level prior to the case study. Although they did not use any metric-driven assessment to evaluate CI maturity, they considered build automation and test writing as key criteria for defining a project's CI maturity.
    }
    \end{minipage}
}
\end{center}

\subsection{RQ2: What benefits does CI monitoring bring?}


In Organization A, one of the most important benefits demonstrated by CI monitoring was its capacity to give an overview of a project's CI stage.  This feature is not usually directly implemented in CI services, as Participant P1 commented: \textit{``It brings precisely the visibility of those aspects that can be considered a problem. As I was talking about this coverage, we already have this information on the \textsc{Sonarqube}. But the other ones are new aspects that bring a more qualitative view of the CI practices we did not have in other tools ... In GitLab, you have a view of the pipeline but do not have an aggregation like in this dashboard, in statistical terms.''}. Monitoring CI practices also proved to be easy to use and low-cost for Organization A.  Participant P1 explained: \textit{`` ... because it is easy to use, it brings value, even if the developer does not use it very often. It has notifications so that developers are always there to know what the values are like. So, in short, it has no major impact on configuration, does not generate extra work, and adds significant value to this aspect of CI practices''}


In Organization B, monitoring CI practices significantly stimulated communication and a desire for improvements. Regarding communication, Participant P4 highlighted: \textit{``yes undoubtedly, we started to hold internal conversations about how to interact more efficiently, we also added revisions for each commit we applied, this was not a common practice...''}. Participant P3 commented on the desire to improve: \textit{``The red images (in dashboard) of the build duration caught my attention. Maybe this sparked my interest in investigating and reducing the dependencies used''}.
Monitoring also helped developers notice previously overlooked problems. Participant P2 described: \textit{``It helped to highlight... The coverage is evident; it was interesting because it showed that there was coverage data, and then suddenly it disappeared. I think it's helping.''}. Participant P3 also said: \textit{``I think it ... pointing out that we are having problems with the code coverage metric. Regarding the build, we already felt that the build took a long time and that the VMs were slow, but not that much, more than 8 hours of build time. We have tests that involve the instantiation of other modules in VMs, ...., but almost 10 hours is a very high value''}.


In Organization C, Participant P5 stated that monitoring CI practices help improve them by emphasizing: \textit{``Specifically, this concern of keeping builds always working, for example. The number of commits, worrying about always committing.''}. It also highlights problems, as noted by P5: \textit{``Having this awareness that the number of commits is important in the project. If there is no significant quantity, we may make commits with a lot of information at once. It helps to cover different aspects''}. Monitoring CI practices also provides an overview of CI stage of a project, as commented by Participant P5: \textit{``It brings more confidence related to some CI needs, for sure. It provided a better view of this process.''}. Finally, it encourages communication: \textit{``We even had a  conversation to discuss these metrics values.''}.


Some benefits were observed across the different projects. One of the most beneficial points highlighted by participants was the reminders and motivation provided to developers. Participant P1 said: \textit{``... it is another thing to be constantly reminded that you have to do it here.''}. Monitoring CI practices is not a priority feature in other tools, as Participant P2 stated: \textit{``I found that the tool consolidates a lot of information in one place that sometimes we have to look for on several \textsc{Sonarqube} screens. It helped a lot. It synthesized it in one place, which I think is cool.''}. The functionality of sending regular alerts to participants was also mentioned as a very important aspect of monitoring. Participant P1 reported: \textit{``Now with the email alert that is sent weekly. It keeps this concern with metrics at the forefront of my mind. Otherwise, if you're focused on other project problems, it's easy to leave it aside''}.

\begin{center}
\fbox{
    \begin{minipage}{24em}
    \textbf{Our findings show that the most important benefits demonstrated by CI monitoring were its ability to provide an overview of the CI stage, highlight CI-related problems, awaken a desire for improvements, improve the quality of CI adoption, and increase communication within the team. CI monitoring also proved to be cost-effective.}
    \end{minipage}
}
\end{center}

\subsection{RQ3: What are the challenges/disadvantages of monitoring CI practices on a daily basis?}


One of the biggest challenges reported in Organization A in using CI practice monitoring daily was not directly related to the monitoring itself, but to acting on the results it provided. Specifically, improving CI maturity, mainly due to time constraints. Monitoring helped highlight the project's lack of testing, but there were obstacles to addressing these issues, as revealed by Participant P1: \textit{``... we end up having to prepare a presentation during the sprint and sometimes I get a little worried about not having time to present what has been completed. If I am going to implement several automated tests, then we end up having to fit the tests into the free time we have in the sprint''}.


When asked about difficulties, Organization B also pointed out that the time dedicated to other development tasks was the main challenge in monitoring and improving CI practices. Participant P2 said: \textit{``... We have very long builds with this build duration in another module, this one has a way to get around it, I just have not had the time yet.''} and Participant P4: \textit{ ``... because the priority is the documentation that needs to be delivered by the end of the year''}. The lack of integration with other CI tools was another problem identified as a difficulty in monitoring Organization B. Participant P3 said:  \textit{``If there was something like a link in the Gitlab interface as \textsc{Sonarqube} has. I could click and be redirected to the tool ... ''} and Participant P4 mentioned:  \textit{``maybe an idea, if you can find some way to add a plugin to Gitlab ... ''}


In Organization C, no relevant challenges in monitoring CI practices were reported. Legacy projects and the internal culture of the project were the main reasons that made improving CI practices, as shown by the tool, challenging. Participant P5 described the difficulty of adopting CI due to the characteristics of the projects, especially legacy systems: \textit{``It's because it uses a legacy system, but it's just a different way of working. The person who developed a large part of the project was a DBA. He worked directly with the tables, and in the end, it was complicated.''} Participant P5 also suggested that better interaction with other tools could facilitate CI monitoring: \textit{``I don't know if the tool already has some type of integration with communication tools, could be a integration with the internal chat that we use for communication ... ''}


During the case study, we also observed additional issues that hinder the appropriate adoption of CI. In Organization A, the pressure for fast deliveries leads to underestimated deadlines, causing certain CI practices, such as writing automated tests, to be frequently postponed or ignored. Delayed testing may detect integration issues only when different parts of the system are combined, resulting to significant delays and increased complexity in remediation. Organization B faces challenges due to a complex development environment, reflected in the instability of the CI environment, which hinders any substantial improvement in the project's CI maturity. Organization C needs to shift its project mindset to realize the full benefits of CI adoption. It was noted that other demands take priority, and developers are not actively encouraged to enhance CI maturity.

\begin{center}
\fbox{
    \begin{minipage}{24em}
    \textbf{ 
    Overall, there were no major challenges directly related to monitoring CI practices. The main challenges reported concerned improving CI practices due to developers’ lack of time caused by other project tasks or the internal culture of the project. Some complaints were made about the tool’s usability and the lack of integration between the monitoring suite (graphical interface) and CI services. Pressure for fast deliveries, complexity and instability in the CI environment, and a lack of emphasis on the importance of CI were identified as problems during monitoring, hindering the improvement of CI maturity.
    }
    \end{minipage}
}
\end{center}

\subsection{RQ4: Are developers interested in using our metrics suite tool?}

\begin{figure*}[ht]
    \centering
    \includegraphics[width=1.0\linewidth]{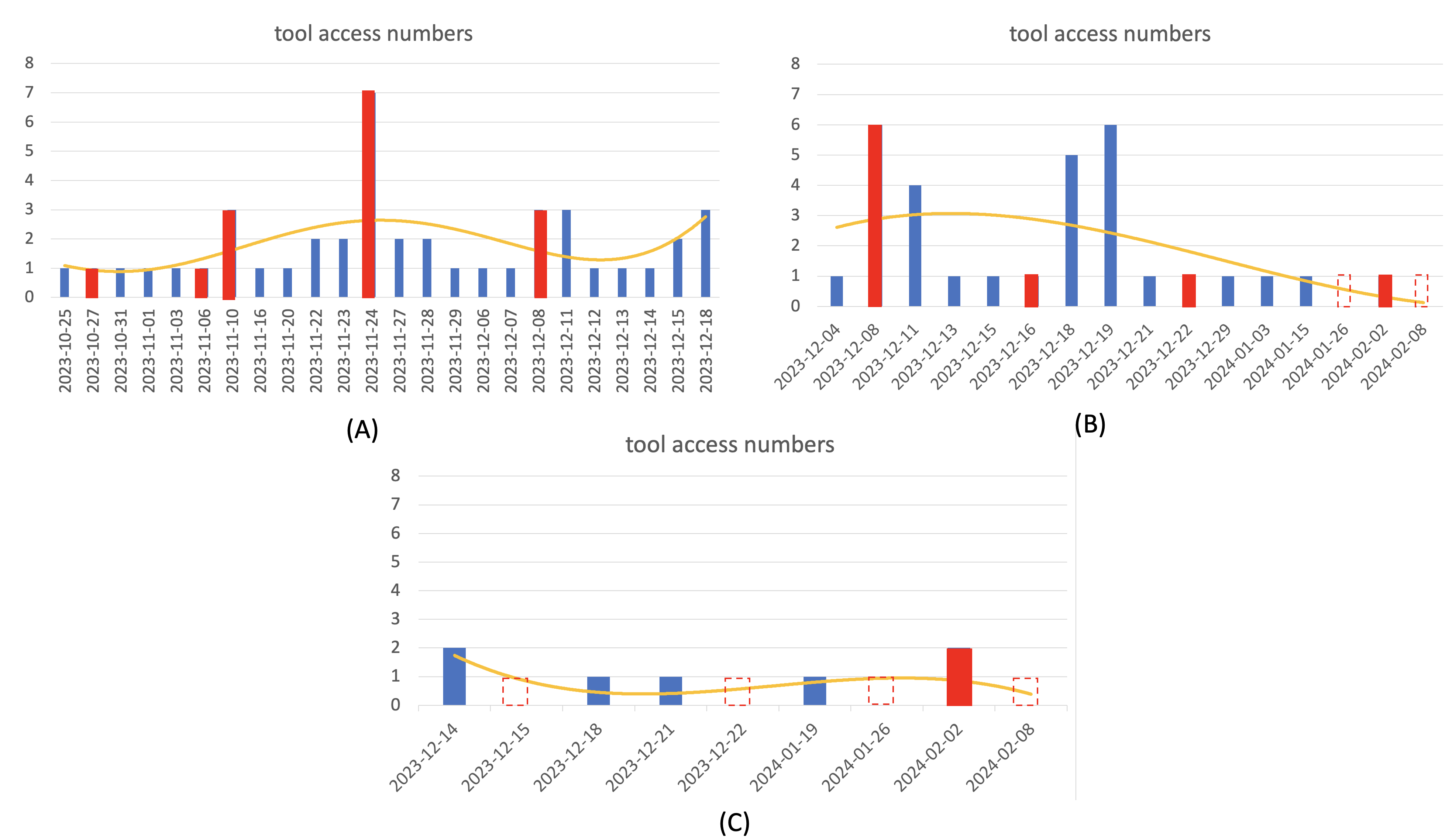}
    \caption{ (A) Number of Access to the monitoring tool of Organization A during the case study.
    (B) Number of Access to the monitoring tool of Organization B during the case study. 
    (C) Number of Access to the monitoring tool of Organization C during the case study.}
    \label{fig:a-number_access}
\end{figure*}

Figure~\ref{fig:a-number_access} shows the tool access data of the three organizations analyzed in the case study. The red bars represent tool access on the days when interviews were conducted. The blue bars represent access on days without interviews. The red crossed-out bars represent days when interviews took place, but no access was recorded in the tool.

In Organization A, during the case study, we observed a total of 43 accesses to the tool dashboard of 2 different developers. In particular, the primary access points coincided with interview dates, as indicated by the red bars on the timeline. These interview-centric peaks suggest increased activity during these periods, potentially linked to collaborative discussions or focused on project evaluations. However, it should be noted that many accesses occurred outside the interview dates, indicating ongoing engagement and usage of the tool beyond the case study evaluation periods. This broader pattern of access underscores the continuous and diverse interactions with the monitoring tool, emphasizing its relevance throughout the case study period.


In Organization B, we observed 31 accesses to the tool dashboard by two different participants. Although three participants took part in the case study, one did not use the tool. We noticed many accesses outside the interview dates (blue bars), indicating interest in monitoring CI practices. There was a sharp decline in access to the monitoring tool after December 22, 2023. This decline can be attributed to two factors: (i) the project going on a holiday break, and (ii) one participant taking vacation during the last two weeks of the case study. Another consideration is that the project was in a phase with few changes. Combined with receiving CI practice values via email, this caused a decrease in interest in accessing the tool interface, as explained by Participant P3: \textit{``I received the email and the issue of code coverage appeared. Since I was already aware of code coverage, I did not pay much attention to the tool because I thought it would be more important to look at the pipeline issue... for the tool to return the collection of the coverage metric correctly''}.


In the last project, Organization C, there was the lowest access to the monitoring tool, with only 7 accesses to the dashboard by one participant. This could be explained by the company's cultural resistance to CI adoption. In addition, the project had a holiday break at the end of the year, which is reflected in the access curve dropping and then rising again when organizational activities resumed. Another observation is that only one participant accessed the monitoring tool during the case study, which could potentially hinder a cultural shift regarding CI within the company.

\begin{center}
\fbox{
    \begin{minipage}{24em}
    \textbf{
    The registration of accesses to the dashboard allows us to conclude that Organizations A and B, which defined themselves as having an INTERMEDIATE CI maturity level, showed greater interest in accessing the tool. Organization C, which has a BASIC CI maturity level, exhibited less interest in using the tool.
    }
    \end{minipage}
}
\end{center}

\subsection{RQ5: How did CI practices evolve during use of our metrics suite tool?}

\begin{figure*}[ht]
    \centering
    \includegraphics[width=\linewidth]{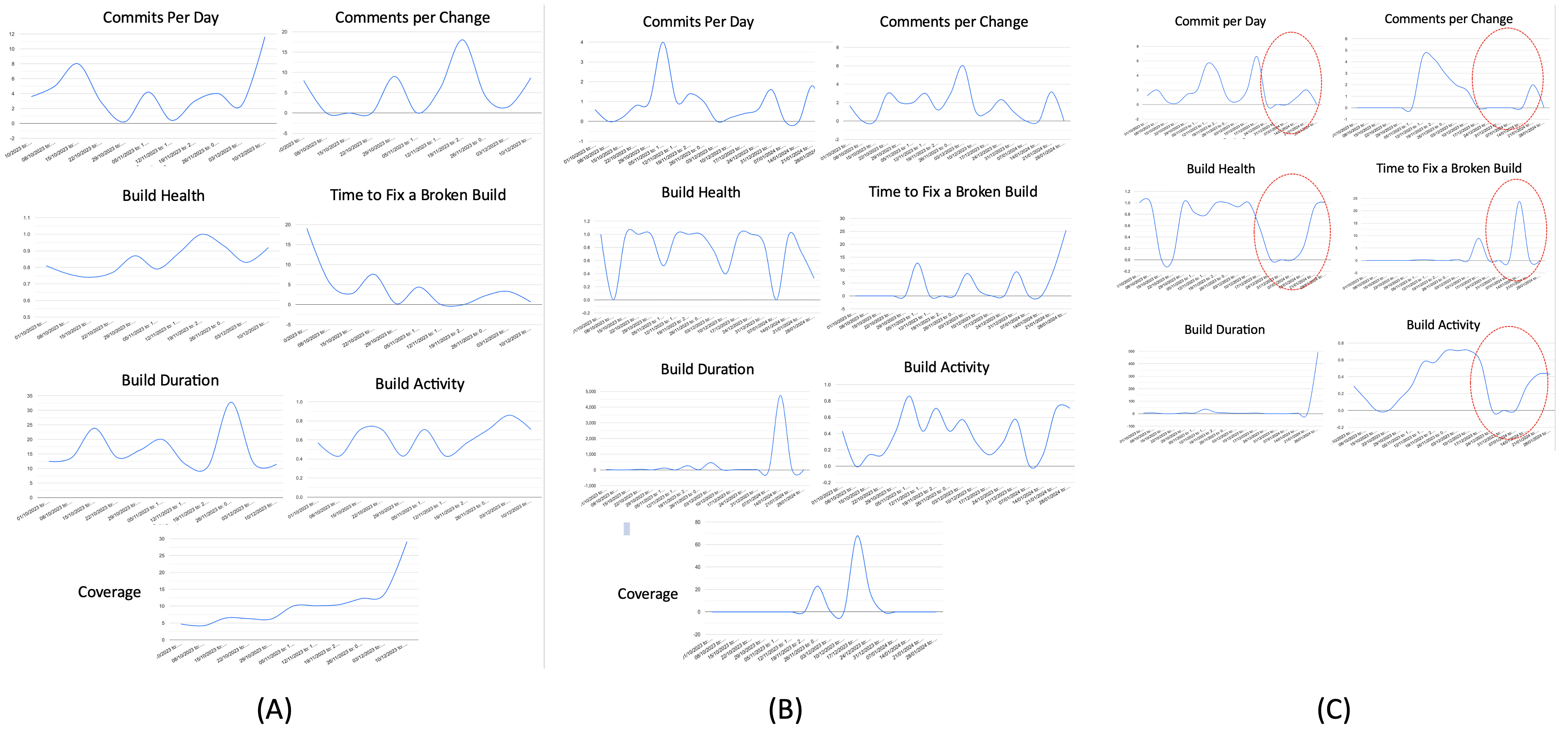}
    \caption{ (A) Organization A CI Practices Evolution during the case study.
    (B) Organization B  CI Practices Evolution during the case study. 
    (C) Organization C CI Practices Evolution during the case study.}
    \label{fig:projects-evolution}
\end{figure*}

Figure~\ref{fig:projects-evolution} highlights the evolution of CI practices for the three organizations during the case study.

In Organization A, since the initiation of our monitoring efforts, all CI practices have demonstrated considerable improvement. Notably, the ``Time to Fix a Broken Build'' experienced a significant reduction, indicating enhanced efficiency in promptly addressing issues. As a consequence, the ``Build Health'' has increased significantly. The improvement in test coverage is equally relevant, and interviews revealed that monitoring practices played a crucial role in driving this positive trend. Participant P1 said: \textit{``I think the evaluation is good for us to have more detailed information, so we can make changes to the project, where necessary. We already knew about the coverage problem from the beginning, but it became very evident. With this information, we began to act to resolve this problem.''}. The ``Commits per Day'' metric dropped in the middle of the case study, followed by a strong recovery. We associate this with the departure of a team member and, subsequently, the entry of new members into the development team. The same team change may have influenced the ``Comments per Change'' metric, which increased during the case study.


Organization B presented a very unstable CI environment due to the complex integration with other applications, which was reflected in the CI metrics. Code coverage mostly remained zero despite the project having automated tests. Build time fluctuated greatly between normal weeks those with extremely high peaks (exceeding 80 hours). Participant P3 reflected on the issues with build time and coverage: \textit{``Yes, we have brought attention to these aspects in terms of build duration and code coverage. We are trying to understand this aspect in meetings, until this morning, the project managers asked me why the tests are taking so long''}. This was a problem they were unable to resolve during the case study. This instability was also reflected in the high variation of ``Build Health'' and  ``Time to Fix a Broken Build'' metrics, as Participant P2 declares: \textit{``Last week we had some problems with infrastructure, we ended up having to do some configurations and such. Because of it, the build health metric started to have a lower value''}. The ``Commits per Day'' metric tended to decrease in the last four weeks of the case study compared to the first four weeks. This trend can be attributed to the conclusion of the development cycle, during which the project shifted its focus towards documentation, as reinforced by participant P4: \textit{``... as everyone is focused on documentation, there shouldn't be any more commits, except something from the README, something very specific like that.''}



In Organization C, the project initially demonstrated good values for CI practices, which was surprising considering the self-declared BASIC level of CI maturity in the initial survey. However, there was a significant degradation in all metrics during the end-of-year holidays break (indicated by the red dotted circle). Despite this, the project experienced a quick recovery afterward in most metrics, such as: ``Build Health'', ``Build Activity'' and ``Time to Fix a Broken Build''. There is evidence that CI monitoring contributed to this recovery, as Participant P5 explained:: \textit{``There was also a period at the end of last year of the build being generated in a broken way, and as soon as we came back, one of the first things we did was fix it... We had already observed this in Gitlab, but the monitoring tool also contributed, in a way, complemented this need to fix it''}

\begin{center}
\fbox{
    \begin{minipage}{24em}
    \textbf{
    Organization A experienced a notable improvement in CI metrics throughout the case study, particularly in coverage metric, where reports indicate that monitoring played a crucial role in identifying and addressing areas needing improvement. In contrast, Organization B faced challenges due to the instability of its CI environment, which impaired significant improvements in the project's CI levels. Finally, in Organization C, the end-of-year holiday break caused a temporary decline in CI metrics, which were quickly improved with the help of monitoring.
    }
    \end{minipage}
}
\end{center}

\subsection{Thematic Analysis}

\begin{figure}[ht]
    \centering
    \includegraphics[width=1\linewidth]{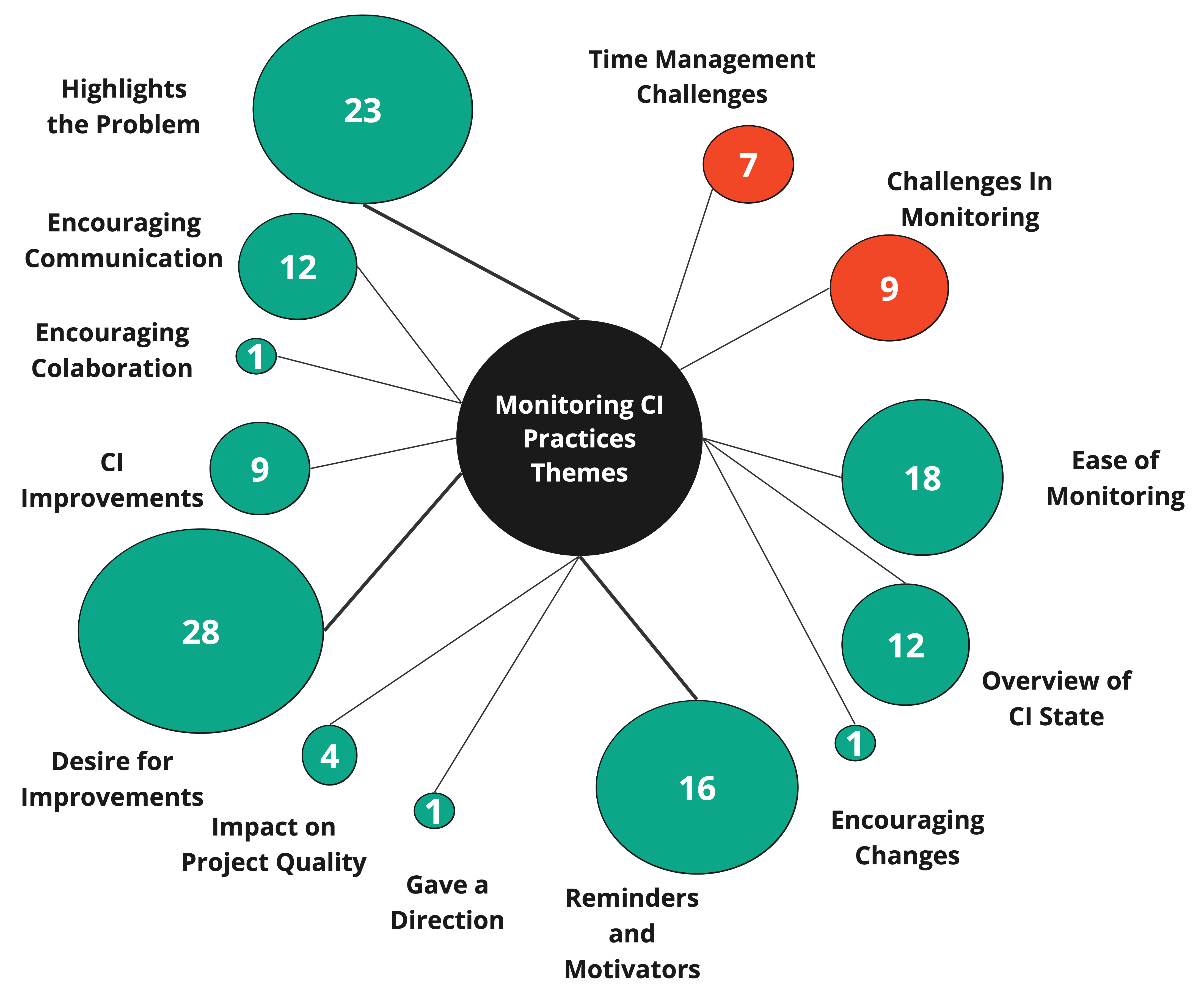}
    \caption{Interviews Themes}
    \label{fig:themes}
\end{figure}

To conclude our case study, we performed a thematic analysis of the participants' interviews. Figure~\ref{fig:themes} presents the main themes that emerged from the interviews about monitoring CI practices. The themes summarize ideas expressed by participants, and the number associated with each theme indicates how frequently it was mentioned. Green nodes represent positive aspects of monitoring CI practices, while red nodes indicate negative aspects highlighted by the interviewees.

The primary positive theme that emerged when monitoring CI practices was the ``Desire for Improvements'' (28 mentions), which illustrates the participant's motivation to enhance certain CI practices in the project (i.e., a desire to improve specific practices in the future). For instance, participant P1 expressed this sentiment by stating: \textit{``We had already been doing this testing issue before, but it certainly reinforced the need to do it as quickly as possible.''}. The second most cited theme was ``Highlights the Problem'' (23 mentions), that the monitoring tool helped the development team discover previously unnoticed issues. Participant P2 articulated this experience, saying:: \textit{``In fact, I found it strange that the coverage dropped, I thought it had already been resolved, there is something there that is being executed and resetting the data.''}. The third most cited theme was ``Ease of Monitoring'' (18 mentions), which reflects how easily developers were able to monitor CI practices using our tool. Participant P4 commented:  \textit{``No, no, it's not expensive. It was very intuitive''}. Finally, ``Reminders and Motivators'' (16 mentions) as another relevant theme, as the tool constantly highlighted issues to developers, motivating them to improve the CI process. As participant P3 remarked: \textit{``... but I think that seeing a very prominent percentage gets attention and helps convince other developers to write tests''}.

Monitoring CI practices also provides an ``Overview of CI State'' for the project (12 mentions), enabling developers to view all CI-related aspects in one place, as participant P2 described: \textit{``... Precisely the tool summarizes several things that perhaps we need to use several tools to get to this information.''}. Additionally, monitoring fosters ``Encouraging Communication'' among team members (12 mentions), as mentioned by participant P5: \textit{``I think it's interesting to know what the practices are like. Just the fact that we talked somewhat superficially about commits would already be a benefit.''}. Moreover, it promotes ``CI Improvements'' (9 mentions), as participant P1 emphasized: \textit{``I am more focused on developing the features that are missing and working on test coverage, which has the most impact for us. I just uploaded some testing issues that will improve the coverage''}. Other themes were mentioned to a lesser extent, such as: ``Impact on Project Quality'' (4 mentions), ``Encouraging Colaboration'' (1 mention),``Gave a Direction'' (1 mention), and ``Encouraging Changes'' (1 mention).

The primary negative theme in monitoring CI practices is ``Challenges in Monitoring''. This emerged from difficulties using the tool or suggestions for improvement indicated by participants during the case study. Most of these issues were related to integrating the tool with other CI tools already used by the team, aiming to facilitate the CI practices monitoring. For example, participant P3 said: \textit{``If there was something like a link in the Gitlab interface as \textsc{Sonarqube} has. I could click and be redirected to the tool ... ''}, participant P5 commented: \textit{``I don't know if the tool already has some type of integration with communication tools, could be a integration with the internal chat that we use for communication ... ''}. Integration with other tools is beyond the initial scope of this study, but it reflects the need for CI tools like GitLab to incorporate monitoring of CI practices directly into their pipelines.

The ``Time Management Challenges'' theme is mainly related to the difficulty teams face in implementing improvements to CI practices due to a lack of dedicated time. This is a common complaint, as developers usually have other responsibilities and cannot focus exclusively on improving CI practices for their projects. Overall, the monitoring tool proved to be easy and intuitive. Monitoring CI practices was not considered burdensome by participants. Participant P4 commented: \textit{``I think the tool is very intuitive, very didactic ...''}. Participant P1 said: \textit{``It has no major impact on configuration, doesn't generate extra work, and adds significant value to this aspect of CI practices.''}. Participant P5 appreciated the alerts sent by the tool: \textit{``I noticed that you can receive it by email. I found the feedback from the emails interesting.''}

\section{Discussion}

In this section, we discuss the implications of our study. During the case study, it was observed that the automation of the build process and the execution of automated tests were the primary factors considered when defining the CI maturity level of projects. However, once CI monitoring was introduced, developers began to emphasize the monitoring process. Monitoring highlighted problems, served as a reminder, and motivated the team to enhance the CI process. They clearly perceived that such improvements could impact overall project quality.

Combining CI practices into a single tool provides an overview of the CI process. It would be beneficial for tools like GitLab to incorporate CI practice monitoring. Such integration would assist in assessing CI maturity. The call for GitLab and similar platforms to support CI monitoring reflects a desire for more seamless and integrated solutions. Direct integration of CI monitoring into popular development platforms streamlines workflows, supports efficient collaboration, and provides real-time insights into the health and effectiveness of CI processes. This approach aligns with the broader industry shift towards DevOps practices, where collaboration between development and operations is tightly integrated.

\textbf{Implications for researchers}. This case study showed that there are still gaps in monitoring development processes. Although CI is already a widespread practice, few studies seek to evaluate its efficient application. There is a field of study exploring new CI practices not evaluated in this study, as well as broader practices such as DevOps. We could also argue that other software engineering practices, in addition to CI, could potentially benefit from monitoring activities. For example, it would be interesting for future studies to monitor security-related metrics and present the results on an organized dashboard.

\textbf{Implications for practitioners}. This study suggests that developers have an interest in the monitoring of CI practices and derive significant benefits from their adoption. Therefore, commercial CI services like GitHub, GitLab, or Jenkins would greatly benefit from incorporating dashboards to monitor development practices such as CI. This addition would add substantial value to these tools and contribute to improving the quality of projects utilizing them.

\section{Threats to Validity}

In this section, we discuss the limitations of our study and how we attempted to mitigate them. 



\textbf{\textit{\underline{Scope:}}} While our case study has provided valuable insights into Continuous Integration practices, its scope was limited to three software development organizations in our immediate geographic area, where we had physical access and permission to install the monitoring suite and collect data. The projects were represented by five participants. As a result, the generalization of our findings is constrained to the project contexts, and the participants may not have a complete understanding of CI across the entire project. We tried to select participants who had knowledge about the project and its CI infrastructure as a way to mitigate this limitation.

\textbf{\textit{\underline{Period of Case Study:}}}  The period during which the study was conducted — eight weeks — may not have been long enough to observe the full effects of any actions taken by the team after becoming aware of the project’s CI maturity. Furthermore, since we were immersed in the project environment, no developer was dedicated exclusively to the case study; they continued with their daily tasks and may not have had time to implement changes in the analyzed project, despite being alerted by the monitoring of CI practices

The case study was carried out from November 2023 until February 2024, coinciding with the holiday season. Therefore, some projects were at the end of the development cycle, with lower demand; final tests or project documentation were carried out, which had a slight impact on some metrics during this period. A break in interviews was held between December 22, 2023, and January 19, 2024, to mitigate this issue.

\textbf{\textit{\underline{Social Desirability Bias:}}} To mitigate social desirability bias, we ensured a private and neutral environment for the interviews and clarified their purpose by emphasizing that all data collected would remain anonymous and be used solely for research purposes.

\textbf{\textit{\underline{Changes in the Project's Team:}}} Organization A experienced changes in its development team during the case study, which may have influenced the evolution of CI practice values reported in RQ5.

\textbf{\textit{\underline{Insufficient Statistical Evidence:}}} Although we observed an improvement in the metrics' values at the end of the case study, future replication studies could help us collect enough data to confirm these hypotheses in a statistical manner.

\textbf{\textit{\underline{Thematic Analysis:}}} In the Thematic Analysis, we acknowledge the potential for bias arising from the authors' subjective interpretations of interview transcriptions. To mitigate researcher bias, the transcript coding was sent to each participant in the case study to validate the generated themes.

\textbf{\textit{\underline{Transcription:}}} To minimize errors during the transcription phase of the case study, we recorded all interviews. Subsequently, we transcribed them using a voice-to-text transcription tool\footnote{https://www.veed.io/tools/audio-to-text}. The transcripts were then verified by the first author, who listened to the interviews and compared them against the automatic tool transcription to ensure accuracy.

\section{Conclusion}

This study presents a case study of three software projects to explore the benefits of monitoring Continuous Integration (CI) practices in real-world settings — a topic often overlooked in both research and industry. To address this gap, the authors developed and deployed a minimally intrusive monitoring suite that allowed teams to monitor their CI practices. Results showed that CI monitoring is valuable for identifying issues, maintaining awareness of CI status, and motivating improvements. Developers expressed a strong interest in integrating monitoring dashboards with tools like GitLab, GitHub, and Jenkins to further streamline their workflows.

Future work should expand CI monitoring to encompass additional DevOps practices, including DORA metrics, to provide a more comprehensive view of the software delivery lifecycle. Integrating deployment and infrastructure metrics would align the monitoring framework with industry best practices and enable a more robust assessment of DevOps effectiveness. A possible direction is the development of a GitHub Action plugin based on the CI monitoring approach, which could be published on the GitHub Marketplace to encourage broader adoption.

\section{Acknowledgments}
This work is partially supported by INES (www.ines.org.br), CNPq grant 465614/2014-0, CAPES grant 88887.136410/2017-00, and FACEPE grants APQ-0399-1.03/17 and PRONEX APQ/0388-1.03/14.

\balance

\end{document}